%% file: plos_template.tex
\documentclass[hidelinks,10pt,letterpaper]{article}

\input{macros}

\usepackage[top=0.85in,left=2.75in,footskip=0.75in]{geometry}
\usepackage{bm}
\usepackage{changepage}

\usepackage[utf8]{inputenc}

\usepackage{textcomp,marvosym}

\usepackage{fixltx2e}

\usepackage{amsmath,amssymb}

\usepackage{cite}

\usepackage{nameref,hyperref}

\usepackage[right]{lineno}

\usepackage{microtype}
\DisableLigatures[f]{encoding = *, family = * }

\usepackage{rotating}
\usepackage{amsmath}

\raggedright
\usepackage[aboveskip=1pt,labelfont=bf,labelsep=period,justification=raggedright,singlelinecheck=off]{caption}

\makeatletter
\renewcommand{\@biblabel}[1]{\quad#1.}
\makeatother

\date{}

\usepackage{lastpage,fancyhdr,graphicx}
\usepackage{booktabs}
 \usepackage{multirow}
\fancyheadoffset[L]{2.25in}
\fancyfootoffset[L]{2.25in}
\begin{document}
\vspace*{0.35in}

% Title must be 150 characters or less
\begin{flushleft}
{\Large The Anatomy of the Three-Point Shot: Spatial Bias, Fractals and the Three-Point Line in the NBA}
\newline
\center{{\tt Working Paper}
% Insert Author names, affiliations and corresponding author email.
\\
Konstantinos Pelechrinis
%Marios Kokkodis\textsuperscript{2},
%,\textpilcrow},
%Name3 Surname\textsuperscript{2,\textcurrency a},
%Name4 Surname\textsuperscript{2,\ddag},
%Name5 Surname\textsuperscript{2,\ddag},
%Name6 Surname\textsuperscript{2,\Yinyang},
%Name7 Surname\textsuperscript{3,*,\Yinyang}
\\
School of Information Sciences\\ University of Pittsburgh\\ Pittsburgh, PA, USA}
%\bf{2} Carroll School of Management, Boston College, Boston, MA, USA
%\\
%\bf{2} Affiliation Dept/Program/Center, Institution Name, City, State, Country
%\\
%\bf{3} Affiliation Dept/Program/Center, Institution Name, City, State, Country
\\

% Insert additional author notes using the symbols described below. Insert symbol callouts after author names as necessary.
% 
% Remove or comment out the author notes below if they aren't used.
%
% Primary Equal Contribution Note
%\Yinyang These authors contributed equally to this work.

% Additional Equal Contribution Note
%\ddag These authors also contributed equally to this work.

%\textcurrency a Insert current address of first author with an address update
% \textcurrency b Insert current address of second author with an address update
% \textcurrency c Insert current address of third author with an address update


% Group/Consortium Author Note
%\textpilcrow Insert Collaborative Author line here

kpele@pitt.edu
\end{flushleft}

\input{texfiles/0-abstract}

%of the review-driven consumer
%\KEYWORDS{deterministic inventory theory; infinite linear programming duality;
%  existence of optimal policies; semi-Markov decision process; cyclic schedule}

%\KEYWORDS{Online Labor Markets, Reputation, Bayesian modeling, Linear Dynamical Systems}

\input{texfiles/1-intro}
\input{texfiles/2-data}
\input{texfiles/3-results}
\input{texfiles/4-discussion}

%\begin{thebibliography}{10}
%\bibitem{bib1}
%Lorem M, Ipsum VE (1990) Rank Correlation Methods. New York: Oxford University Press, 5th edition.

%\end{thebibliography}

\section*{Supporting Information}

\begin{figure}[h]
\begin{center}
%\vspace{-0.2in}
\includegraphics[scale=0.27]{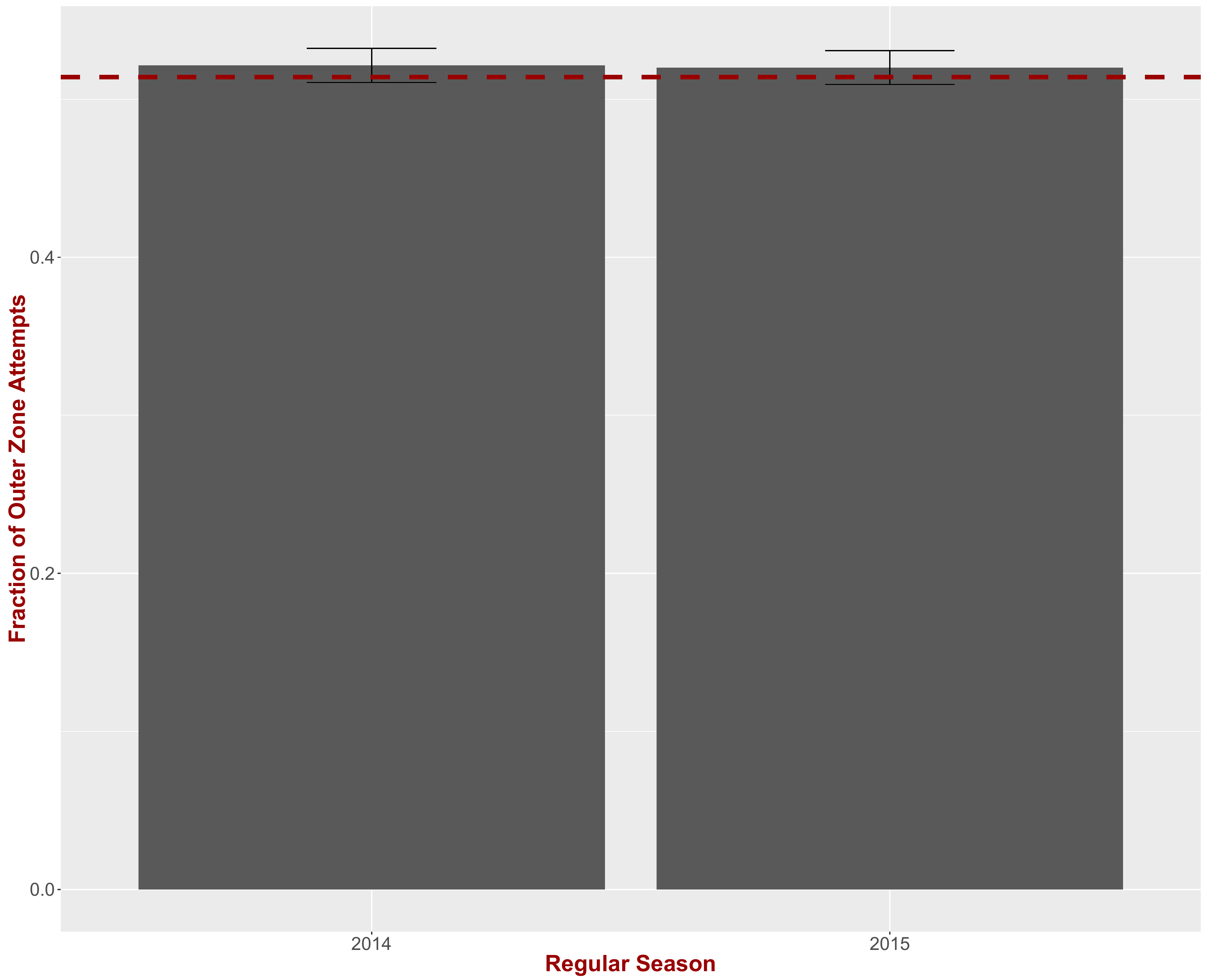}
\caption{When examining the one-feet zones around the 17-feet distance, we do not observe any spatial bias.}
\label{fig:2pt-fractions-control}
\end{center}
%\vspace{-0.3in}
\end{figure}

\paragraph*{S1 Appendix.}
\label{S1_Appendix}
{\bf Control Case for 2PT shots.} We examine a control case, that is, a two-feet zone around the 17-feet distance arc (this is a mid-range shot, that regardless of the location in this area the shot is a two-point shot) in order to increase our confidence that the spatial bias observed around the three-point line is not something that is observed even into mid-range two-point shots.  
The results are presented in Figure \ref{fig:2pt-fractions-control}, where the baseline fraction is also presented (given the arc shape and the radius of the zone this is 51.4\%).  
As we can see in this case there is no spatial bias; the 95\% confidence intervals of the observed fractions includes the baseline fraction!  
The players exploit {\em equally} the examined areas, in contrast to the zones around the three-point line. 

\end{document}

%% file: macros.tex
\usepackage{xcolor}
\usepackage{mdframed}
\usepackage{bbm}
\usepackage{adjustbox}
\usepackage{multirow}
\usepackage{rotating}
 \usepackage{array,multirow,graphicx}

\newmdtheoremenv{conj}{Conjecture}
\newmdtheoremenv{lemma}{Lemma}

\newcommand{\fracdim}{{D}}
\newcommand{\hopcount}{{r}}

\newcommand{\equity}{\mathcal{E}}

\newcommand{\indicator}{\mathbbm{1}_{3PT}}

%% file: texfiles/0-abstract.tex
Even though it might have taken some time, the three-point line ultimately changed the way the game is played as evidenced by the increase in the three-point shot attempts over the years.  
However, during the last few years we have experienced record-breaking seasons in terms of both three-point attempts and field goals made.  
This brings back to the surface questions such as ``What is the rationale behind the three-point line?'', ``Is the three-point shot distance appropriate?'' and many more similar questions. 
In this work, even though we do not take a stand against the three-point line, we provide evidence that challenge its distance.  
In particular, we analyze shot charts and we identify a statistically significant discontinuity in the shot attempts between 1-feet zones just inside and outside the three-point line.  
In addition we introduce a metric inspired by fractal theory to quantify this bias and our results clearly indicate that the space dimensionality in these areas of the court is not fully exploited.  
By further examining the field goal percentages in the zones considered, we do not identify a similar discontinuity, i.e., the field goal percentage just inside the three-point line and just outside the line are statistically identical.  
Therefore, even though the shooting behavior of the teams appears to be {\em rational}, it raises important questions about the rationality of the three-point line itself.

%% file: texfiles/1-intro.tex
\section{Introduction}
\label{sec:intro}

During last two-seasons the Golden State Warriors have introduced a game plan that is heavily relying on three-point shots.  
Fans are split on this with some finding it fascinating - who did not enjoy seeing Klay Thompson set a new record for the number of three-points shots made for a single game agains Oklahoma City Thunders - while other boring and even hurtful for the game\footnote{\small{\url{http://www.usatoday.com/story/sports/nba/2015/12/27/mark-jackson-stephen-curry-comparing-three-point-shooting-over-the-years/77948716/}}}.  
Despite where one lays on the spectrum of preference for this type of game plan, the hard fact is that teams rely on three-point shooting more than ever before. 

However, have the teams that rely heavily on three-point shots simply beaten the game by identifying inefficiencies in its current form? 
Can these inefficiencies be eliminated? 
Will this be a long term solution or just a short term remedy? 
These are some of the questions that our analysis in this paper aims into answering.  

\begin{figure}[h]
\begin{center}
%\vspace{-0.2in}
\includegraphics[scale=0.27]{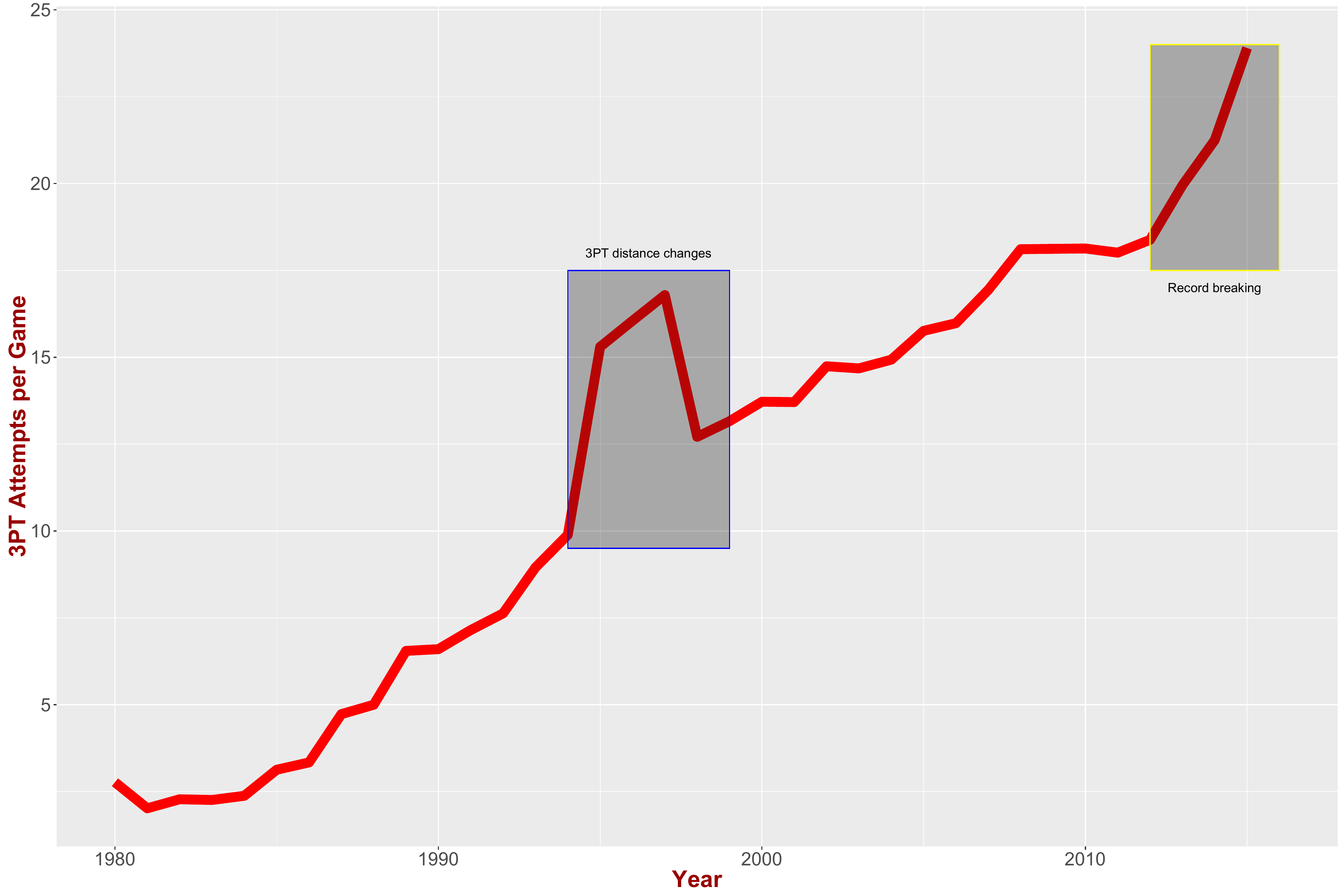}
\caption{There is a stable increasing trend in the three-point attempts per game the last 3 seasons.  }
\label{fig:3pta-year}
\end{center}
%\vspace{-0.3in}
\end{figure}

Figure \ref{fig:3pta-year} presents the three-point attempts per game since the 1980-81 season.  
As we can see the last two-years the increasing trend in shooting is comparable only for the increase observed during 1995 when the distance to the three-point line was shortened.  
The increase in three-point attempts is not necessarily bad from the onset.  
For example, teams might be taking calculated risks when focusing on the outside threat.  
In fact, as our analysis will reveal, this behavior is a response to {\em game design inefficiencies} and is completely rational from a coaching perspective.   

However, simply comparing aggregate numbers cannot reveal a lot about the behavior of teams with regards to the three-point shot.  
In this work, we analyze shot charts from the last two NBA regular seasons.  
Shot charts have been central to the analysis of the game due to the richness of information included in them \cite{reich06,goldsberry12,Chang14}.  
Our goal is to describe and understand the behavior of teams around the three-point line.  
Our analysis reveals that there is a clear bias on shot attempts just outside the three-point line as compared to those just inside the line.  
Furthermore, the field goal percentage for these two zones is statistically identical. 
This marks a clear inefficiency of the game, since two shots with an apparent equal difficulty are rewarded differently!  
We also use ideas borrowed from fractal theory, and in particular the notion of fractal dimensionality, to quantify the spatial bias of the three-point attempts.  

The rest of the paper is organized as follows.  
In Section \ref{sec:data} we describe our data and required background for our analysis, while in Section \ref{sec:results} we present the results of our analysis.  
Finally, \ref{sec:discussion} discusses the implications of our findings and concludes our study. 

%% file: texfiles/2-data.tex
\section{Materials and Methods}
\label{sec:data}

In this section we will briefly describe the dataset we collected and used, while we will also provide some necessary background for our analysis.  

\begin{figure}[h]
\begin{center}
%\vspace{-0.2in}
\includegraphics[scale=0.27]{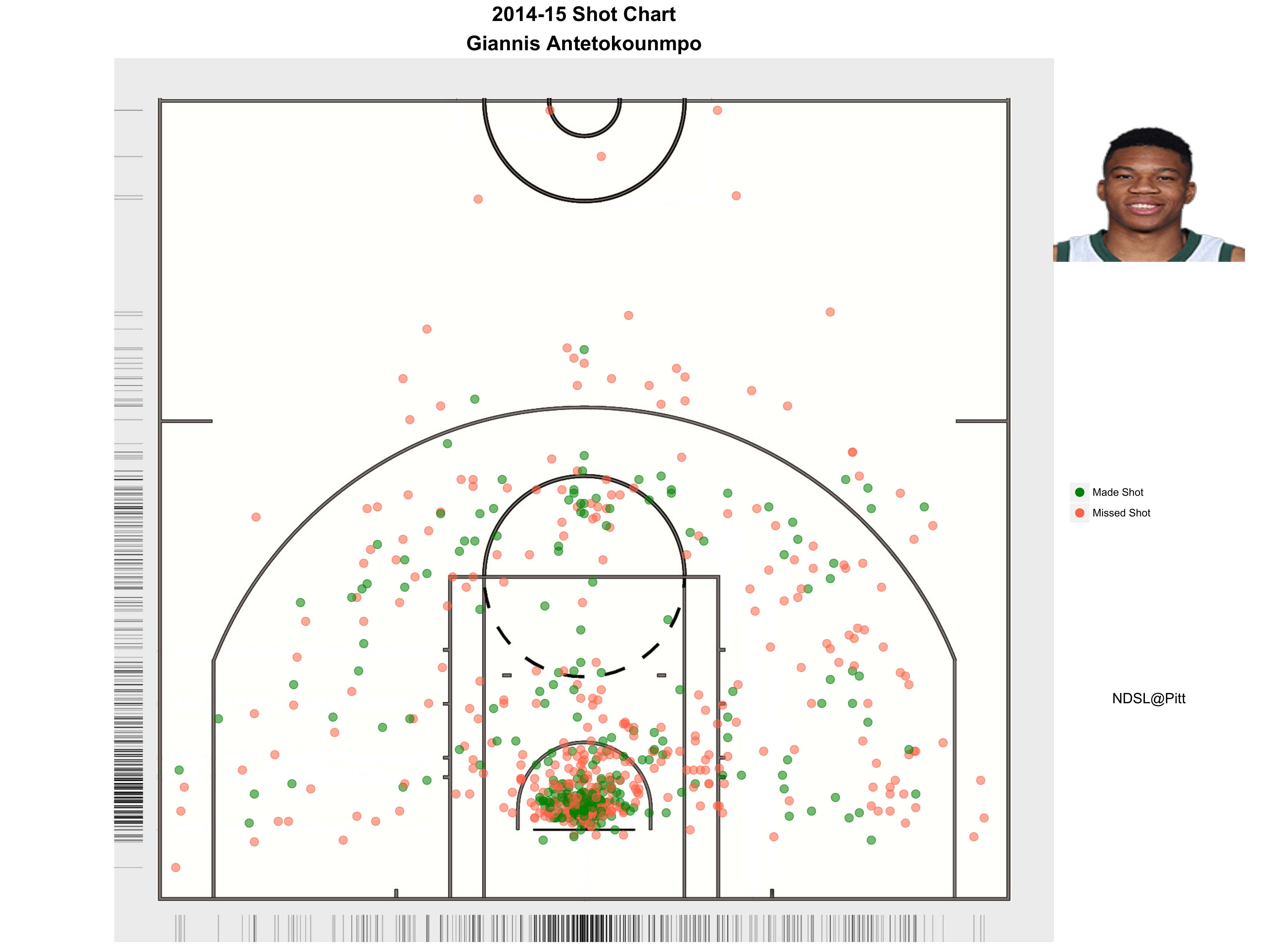}
\caption{An example of a shot chart for Giannis Antetokoumpo.  Shots made are marked with green color, while missed shot are labeled with red.}
\label{fig:giannis}
\end{center}
%\vspace{-0.3in}
\end{figure}

{\bf Dataset: }
Using NBA's API we collected detailed shot charts\footnote{Data and scripts can be made available upon request.} for each player during the last two regular seasons, namely, 2014-15 and 2015-16.  
Each data point corresponds to a shot taken (made or missed) and includes detailed information in the following tuple format: 
{\tt <Game\_ID, Game\_Event\_ID, Player\_ID, Player\_Name, Team\_ID, Team\_Name, Period, Minutes\_Remaining, Seconds\_Remaining, Event\_Type, Action\_Type, Shot\_Type, Shot\_Zone, Shot\_Distance, Location\_X, Location\_Y, Shot\_Made\_Flag>}. 
For our study particularly important is the information about the {\tt Shot\_Distance}, the location of the shot ({\tt Location\_X} and {\tt Location\_Y}) and the {\tt Shot\_Zone} (i.e., whether the shot were from the corners or the crest).  
Figure \ref{fig:giannis} depicts an example of a color-coded shot chart for Giannis Antetokoumpo for the 2014-15 regular season.  
As we can see, shot chart datasets provide a very detailed view of the shots taken in the league and are the appropriate source for the purposes of our study.  

{\bf Fractal Dimension: }
Fractal theory can quantify the dimensionality structure of complex geometric objects beyond their pure topological aspects.  
In contrast to topological dimensions, fractal dimensions can take non-integer values allowing us for a more detailed description of the space that the object of interest fills \cite{schroeder91}. 
While there are various definitions for fractal dimension, the most appropriate for spatial data is the definition of the fractal correlation dimension $\fracdim_2$ on a cloud of points $\mathcal{S}$.  
In particular, with $C(\hopcount)$ being the fraction of pairs of points from $\mathcal{S}$ that have distance smaller or equal to $\hopcount$, $\mathcal{S}$ behaves like a fractal with intrinsic fractal dimension $\fracdim_2$ in the range of scales $r_1$ to $r_2$ iff: 

\begin{equation}
C(\hopcount)\propto r^{\fracdim_2}~~~~~~ r1 \le \hopcount \le r2
%\fracdim_2 = \lim_{r\rightarrow 0}\dfrac{\log C(r)}{\log r}
\label{eq:d2}
\end{equation}
%where $C(r)$ is the fraction of pairs of points from $\mathcal{S}$ that have distance smaller or equal to $r$.  
%In other words, the point-set $\mathcal{S}$ is behaving like a fractal with intrinsic fractal dimension $\fracdim_2$ in the range of scales $r_1$ to $r_2$ if $C\propto r^{\fracdim_2}$ for this range of scales.  
An infinitely complicated set $\mathcal{S}$ would exhibit the above scaling over all the possible ranges of $r$.  
However, real objects are finite and hence, Equation (\ref{eq:d2}) holds only over a specific range of scales.  
For example, a cloud of points uniformly distributed in the unit square, has intrinsic dimension $\fracdim_2 = 2$, for the range of scales $[r_{min},1]$, where $r_{min}$ is the smallest distance among the pairs of $\mathcal{S}$.  

The correlation dimension $\fracdim_2$ can be used to describe shot charts.  
This can either include full shot charts, or subsets of them (i.e., over a particular space of interest).  
In what follows we will use the fractal dimension $\fracdim_2$, to quantify the spatial bias of three-point shots and the inefficiency related with the use of the space around the three-point line. 

%% file: texfiles/3-results.tex
\section{Results}
\label{sec:results}

In this section we will present our results.  
In particular, we will start by analyzing the shots taken around the three-point line both in aggregate and based on their spatial distribution. 
We will then explore whether there is a way to remove the inefficiency identified by rethinking the three-point line.

\subsection{Spatial Bias}
\label{sec:bias}

We begin by focusing on an one-feet zone just inside the three-point line and an one-feet zone just outside the three-point line.  
The distance of the three-point line is different at the corners (22 feet) and the crest (23.9 feet) and hence, we analyze these cases separately. 
Furthermore, the shape of the three-point line is different in the corners (straight line) and the crest (arc).  
This impacts the baselines that we will use for comparison as we elaborate on in what follows.

\begin{figure}[h]
\begin{center}
%\hspace{-0.2in}
\includegraphics[scale=0.25]{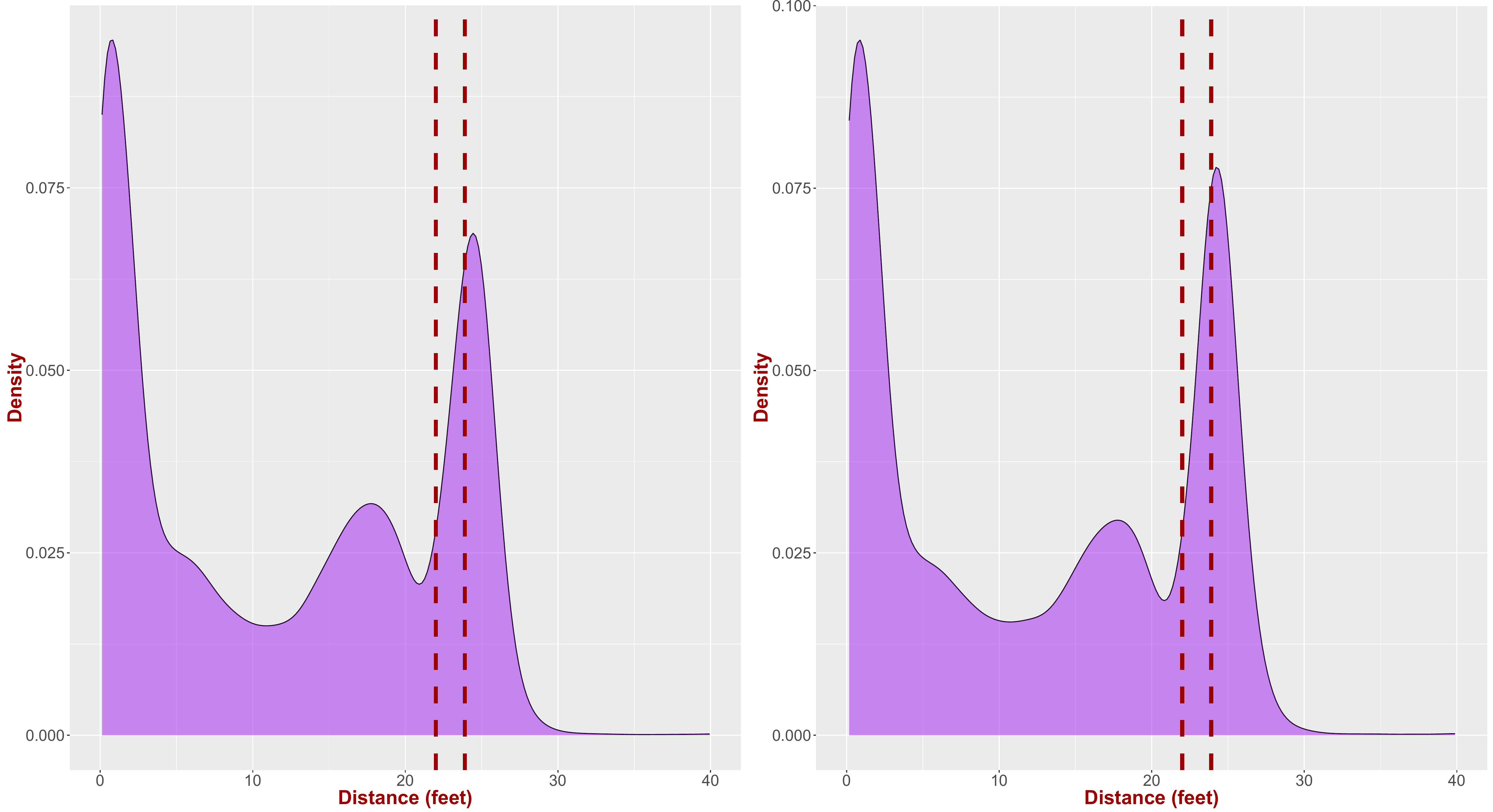}
\caption{The probability density function of the shot distance is fairly stable between the two seasons examined (2014-15 on the left and 2015-16 on the right).}
\label{fig:pdf_dist}
\end{center}
%\vspace{-0.3in}
\end{figure}

We begin by analyzing the distribution of the shots based on their distance from the basket.  
Figure \ref{fig:pdf_dist} depicts the probability density function for the two seasons we examined.  
As we can see there are various local minima and maxima.  
Players tend to take shots mostly around the basket, while the next largest local maximum is observed around the 24 feet distance.  
While these density plots gives us a basic idea of how players take their shots based on distance it cannot really capture a lot of spatial granularity. 
One of the problem is that given that there are two different distances for the three-point line (marked with the dashed lines on the figure), we cannot know from these results whether the shots between the range of 22 and 23.9 feet are three or two point attempts.  

\begin{figure}[h]
\begin{center}
%\vspace{-0.2in}
\includegraphics[scale=0.27]{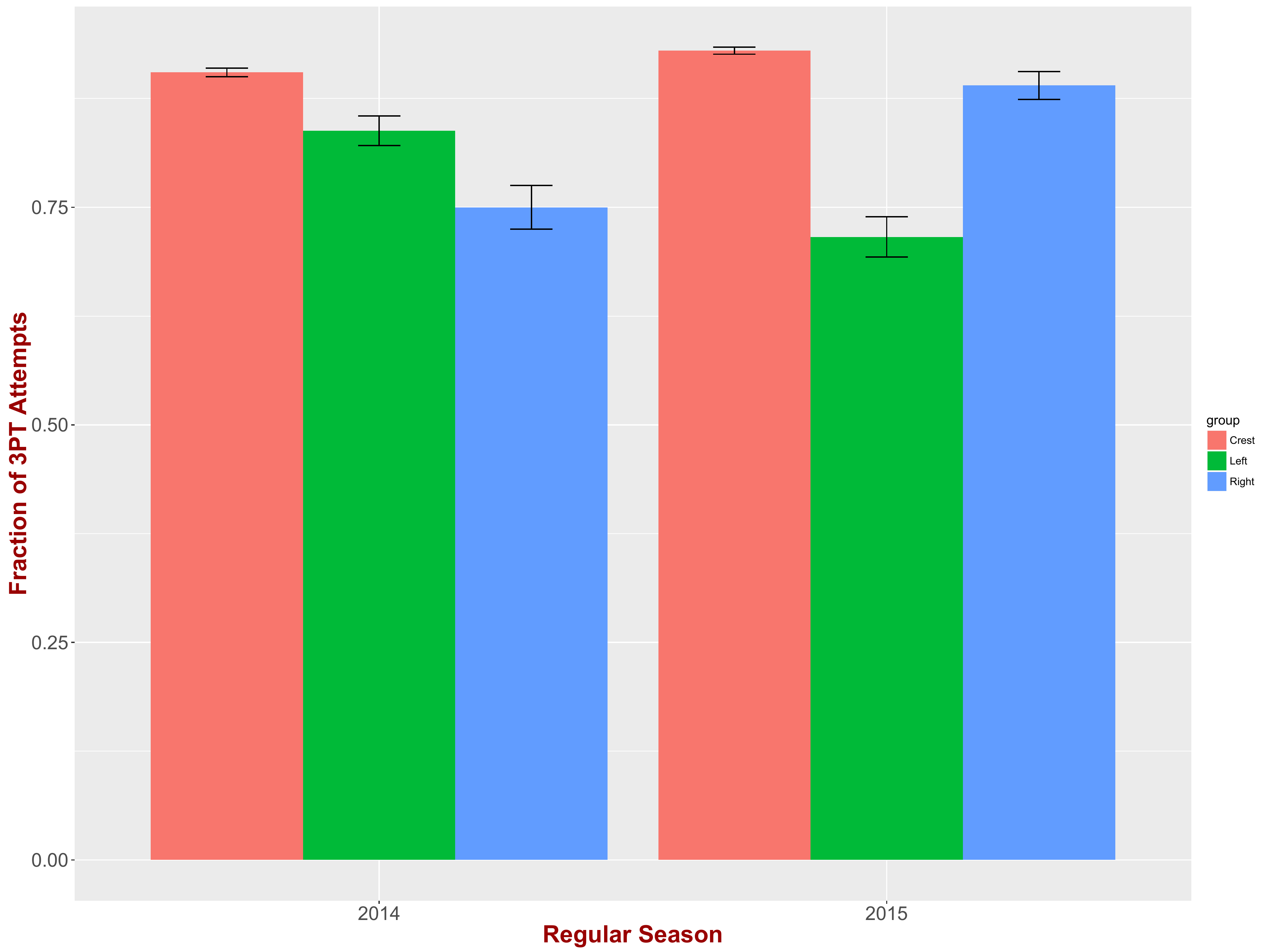}
\caption{The fraction of three-point shots taken in the zone examined is significantly higher to what one might have expected if there was no spatial-bias around the three-point line.}
\label{fig:3pt-fractions}
\end{center}
%\vspace{-0.3in}
\end{figure}

Given the above problem we analyze the fraction of shots within each two-feet zone around the three-point area that were three-point attempts.  
The thesis behind this calculation is that if there was not any spatial bias, the shots would be {\em evenly} allocated within the zone just inside the line and that just outside the line.  
In the case of the corner threes, where the shape of the three-point mark is a straight line, the expected baseline is 50\%, i.e., one should expect in the absence of any spatial bias, 50\% of the shots to be just outside the line and 50\% just inside.  
For the crest threes, where the shape of the three-point line is an arc, the one-feet area just inside the three-point line covers 52\% of the total area examined and hence, one would expect 52\% of the shots be just outside the line.  
Figure \ref{fig:3pt-fractions} presents our results where we can see that in all field locations and for both seasons, the fraction of three points taken in the zone examined are statistically significant\footnote{We use a confidence level of $\alpha=0.05$ unless otherwise specified.} higher than the expected one if no spatial bias was present.  
The most startling result is that the difference between the actual fraction and the baseline is between 35-40\%.  
In  \nameref{S1_Appendix} we further present a control case, which aims at examining whether the observed bias is more inherent to the game and not really related with the three-point line.  
The results indicate that this bias is not present in mid-range shots.  

\begin{figure}[h]
\begin{center}
%\vspace{-0.2in}
\includegraphics[scale=0.27]{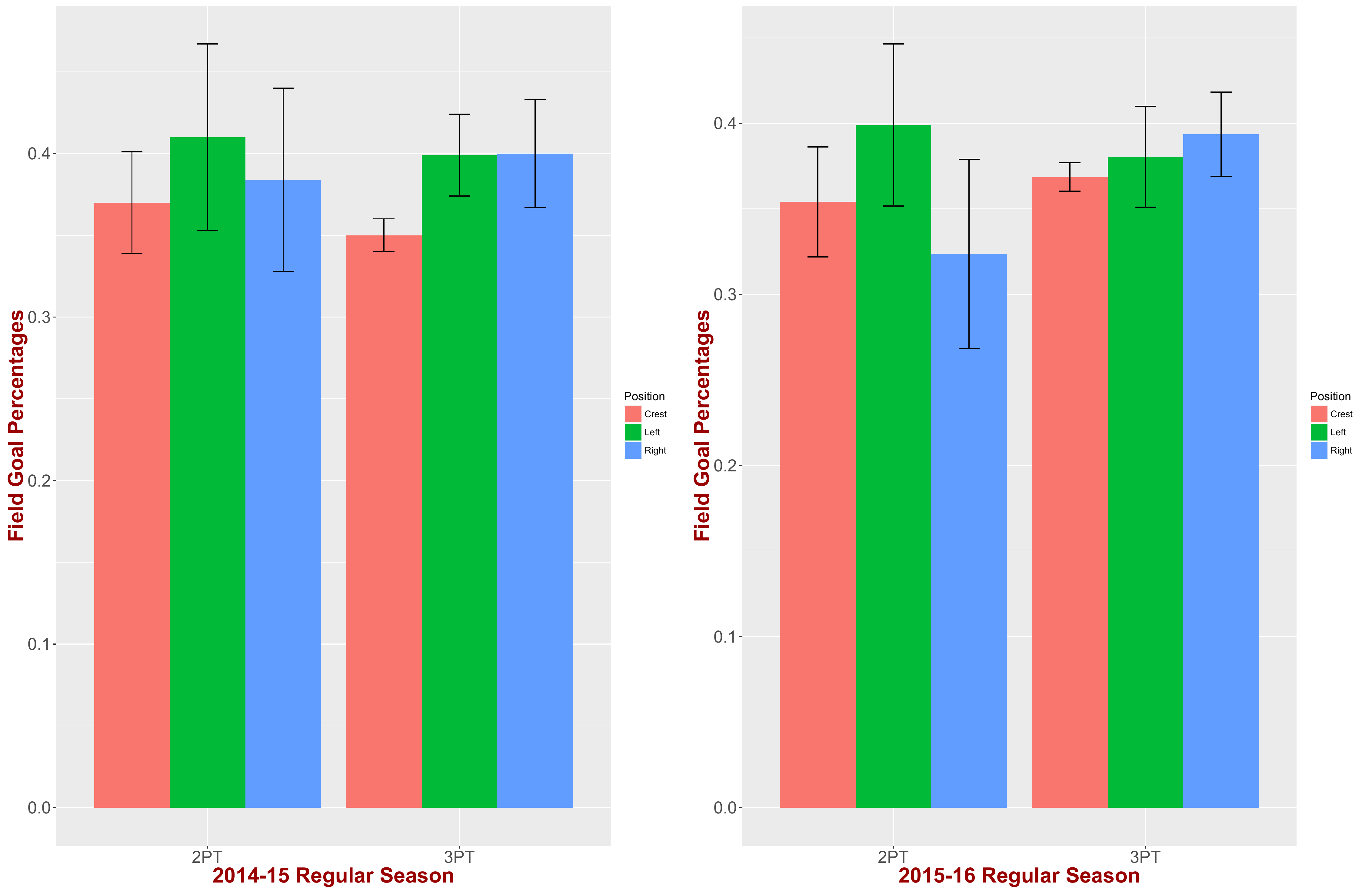}
\caption{There is no statistically significant difference between the field goal percentage just outside and just inside the three-point line.}
\label{fig:fgp}
\end{center}
%\vspace{-0.3in}
\end{figure}

We then examine the field goal percentages for the same zones around the three point line.  
In particular, we want to examine whether the discontinuity observed at the field goal attempts between the two zones is accompanied with a similar discontinuity at the field goal percentage.  
The results are presented in Figure \ref{fig:fgp}, where we can see that when comparing the two and three-point field goal percentages for the same position, there is not any statistically significant difference.  
Simply put, two shots with the same level of difficulty can potentially be rewarded with different number of points.  
This essentially means that players behave exactly as a rational agent would, i.e., bias their shots outside the three-point line.

What the above analysis reveals is that there is an inefficiency in the game, since a shot, say shot $x$, that is equally as hard as shot $y$ might be rewarded with 33\% more points.  
The area just inside the three-point line appears to provide less {\em court equity} $\equity(x,y)$ to the teams/players and thus, the space is not utilized efficiently.  
If we consider the court equity $\equity(x,y)$ to be proportional to the expected points gained through a shot from position $(x,y)$, we would have: 

\begin{equation}
\equity(x,y) = FGP(x,y)\cdot [(2\cdot(1-\indicator(x,y))) + (3\cdot \indicator(x,y))]
\label{eq:equity}
\end{equation}
where $FGP(x,y)$ is the field goal percentage for location $(x,y)$, while $\indicator$ is an indicator function on whether the shot taken is a three-point attempt.
We can see that in order for the equity of the two zones examined to be equal, and therefore, to not have inefficient use of the space, $FPG_{OUT} = 0.66 \cdot FPG_{IN}$, where $OUT$ ($IN$) represents the one-feet area just outside (inside) the three-point line. 

In the following, we quantify this inefficiency in utilizing the space using the notion of fractal dimensionality introduced in Section \ref{sec:data}.

\subsection{Spatial Inefficiency}
\label{sec:fractals}

One way to quantify the aforementioned inefficiency is through estimating the fractal dimensionality of the point set comprised of the shot locations that fall within the zones examined.  
We can then compare it with the dimensionality expected if the shots were taken without any bias, i.e., distributed uniformly over the area examined. 
The benefit of using the fractal dimensionality lays on the fact that it is not restricted to integer values.  
In particular, the topological dimensionality of the basketball court is equal to 2.  
However, the players do not exploit the different court locations equally and therefore, the {\em effective} dimensionality is reduced. 

We begin by examining the actual dimensionality observed in the 3 zones we considered in our analysis - left corner, right corner and the crest.  
Given that both seasons exhibit similar dimensionality we present the results for both seasons together.  
Table \ref{tab:frac_dim} presents our results. 
Along with the computed dimensionality we present the dimensionality that should have been expected if there was not any spatial bias.  
This was computing by reshuffling the shots taken in the area uniformly at random within the area of interest.

\begin{table}[h]
  \begin{center}
    \begin{tabular}{l|c|c}
    \toprule
\bf Position & $\fracdim_2$ & \bf Theoretical dimensionality \\
      \midrule
%Intercept &  -3.250511$^{***}$\\
Left Corner & 0.67 & [0.96, 0.99]\\
Right Corner & 0.7 & [0.97, 0.99]\\
Crest & 1.17 & [1.74, 1.87]\\
\bottomrule
    \end{tabular}
    \vspace{0.2cm}
       \caption{The dimensionality observed on the court locations around the three-point line is much smaller as compared to the one expected if there was no spatial bias in shot taking.}
    \label{tab:frac_dim}
  \end{center}
\end{table}

\begin{figure}[h]
\begin{center}
%\vspace{-0.2in}
\includegraphics[scale=0.5]{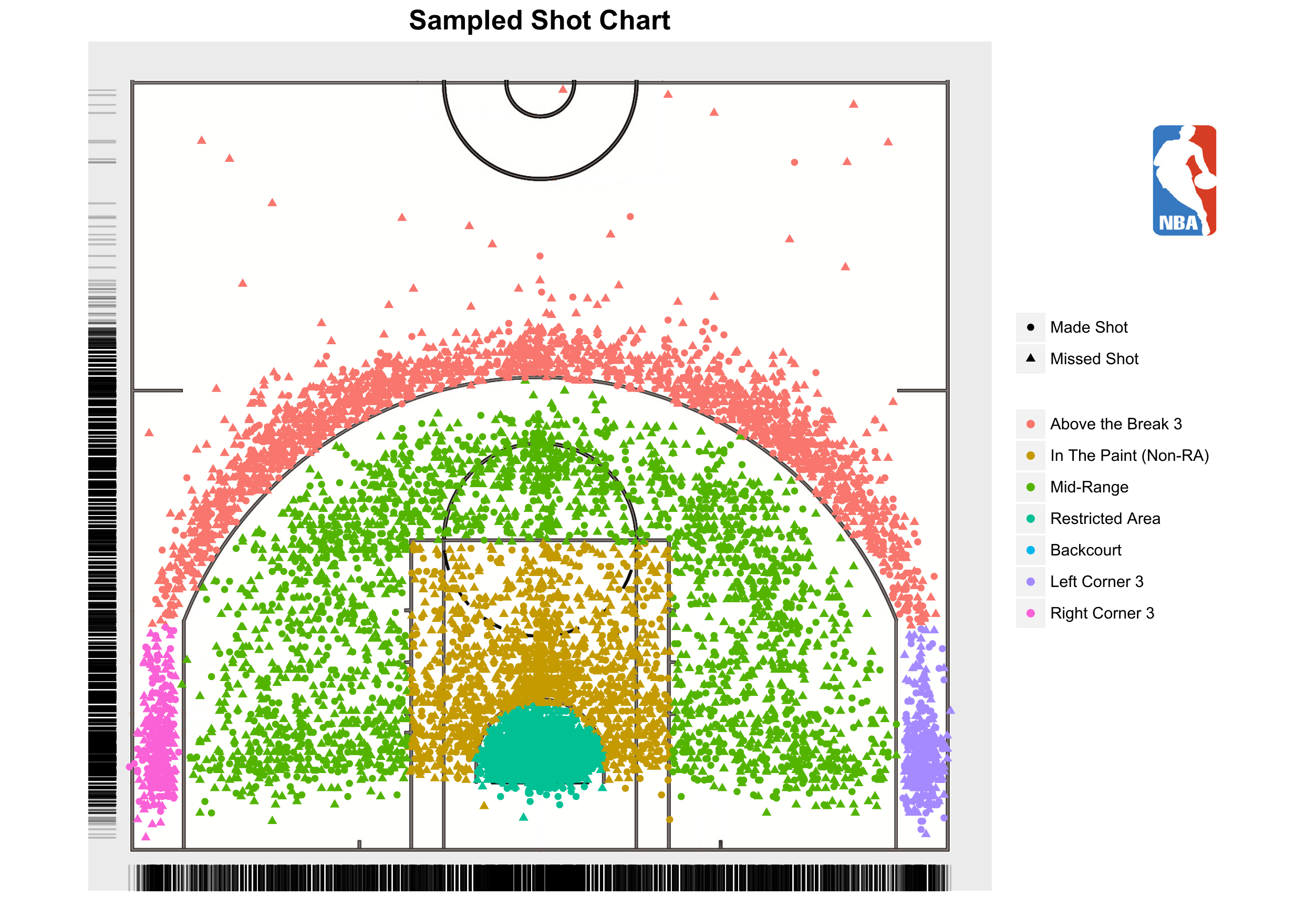}
\caption{There is an approximately 30\% reduction in the dimensionality of the court floor utilized around the three-point line.}
\label{fig:shot-charts}
\end{center}
\vspace{-0.3in}
\end{figure}

As we see the shooting behavior of the players appears to be single-dimensional even in the crest three-point area, where the expected dimensionality is almost equal to the topological one.  
For the corner areas, the theoretical/expected dimensionality is much smaller since the complexity of the area examined is significantly smaller (practically it is a narrow {\em straight line}).  
Nevertheless, the actual dimensionality is still significantly smaller.  
Overall, in both the corners and the crest, there is an approximately 30\% reduction in the dimensionality.
Figure \ref{fig:shot-charts} depicts a uniformly sampled (for better visualization) shot chart with the shots color-coded based on the different courts areas within where they were taken.  
As one can observe, there is a significant absence of shots just inside the three-point line as compared to the density of the shots taken just outside the three-point line.  

\subsection{Data-driven Recommendations}
\label{sec:recommendation}

Despite the above analysis the question still remaining is how can one eliminate this observed inefficiency in the court utilization. 
In our opinion, from a pure equity perspective, the three-point line should mark a location where there is a statistically significant discontinuity in the field goal percentage of the shots taken just inside the line and just outside the line.  
One might argue that the equity of every court location should be constant, i.e., that the reward should be tied with the difficulty of making the shot.  
Of course, this might cause other problems that we discuss in Section \ref{sec:discussion}.  

\begin{figure}[ht]
\begin{center}
\vspace{-0.3in}
\includegraphics[scale=0.35,angle=270]{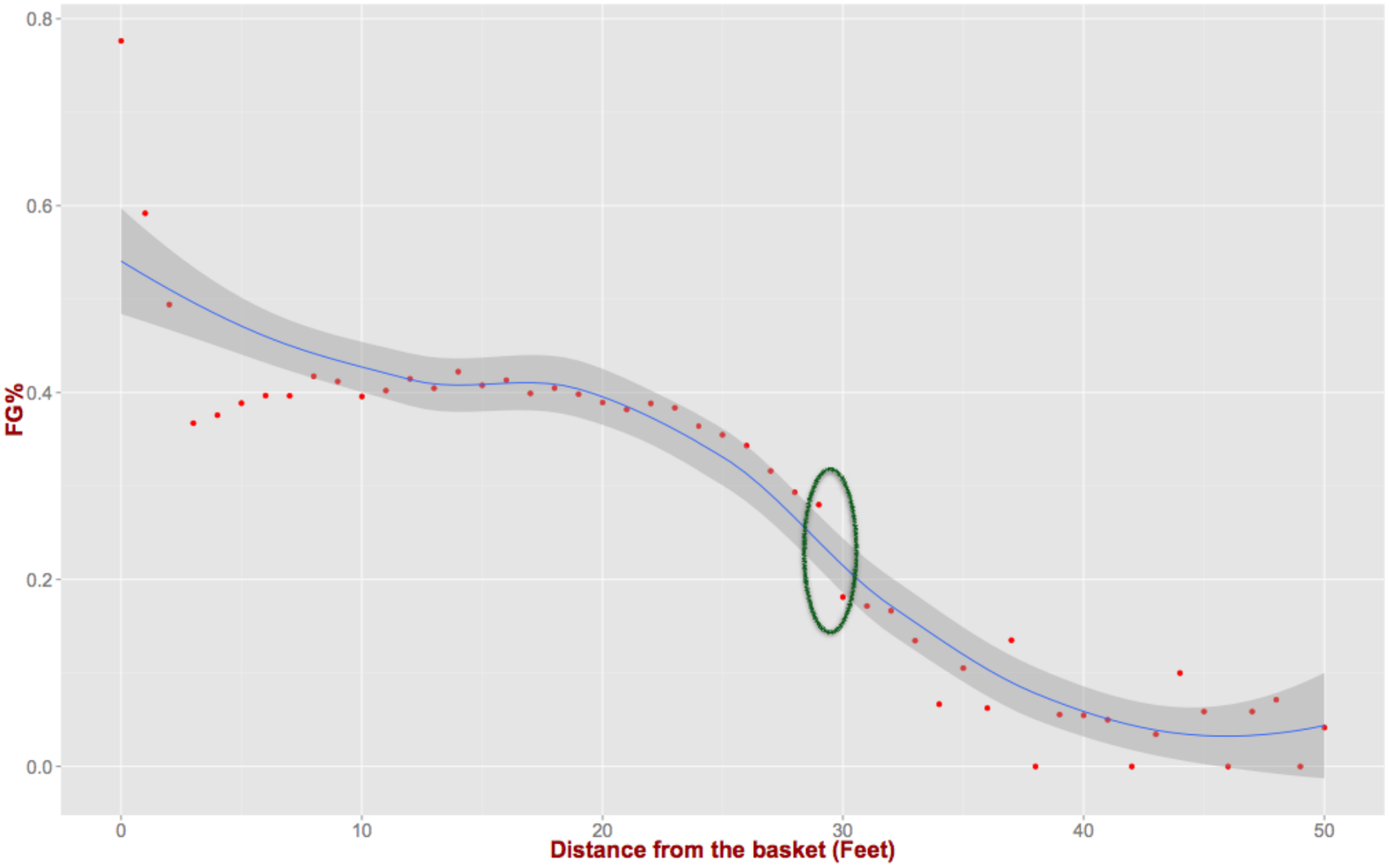}
\caption{There is an approximately 30\% reduction in the dimensionality of the court floor utilized around the three-point line.}
\label{fig:fgp-dist}
\end{center}
\vspace{-0.35in}
\end{figure}

To see whether there is any distance from the basket that a significant discontinuity exists for the field goal percentage, we compute $FPG$ as a function of distance.  
Figure \ref{fig:fgp-dist} presents our results.  
As we can see overall there is a smooth transition - and not statistically significant change - of the $FGP$ over the different distances.  
Of course, there are significant discontinuities for the very small distances to the basket, which however for obvious reasons one cannot consider being the three-point line.  
The only distance where there is a discontinuity (p-value $<$ 0.1) in the field goal percentage is 30 feet.  
The observed discontinuity (9\% change) still provides more equity at the three-point shot, however, the differential is not reduced.  
Of course, there are many things to consider before changing the three-point line.  
In particular, players are expected to adopt to the new distance and this discontinuity might (quickly or slowly) disappear.  
Furthermore, a 30-feet three-point line distance means that the court dimensions need to change, in order for the corner threes be adjusted as well. 
We briefly discuss these topics in the following section.

%% file: texfiles/4-discussion.tex
\section{Discussion}
\label{sec:discussion}

The objective of our work is twofold and should be treated as such; (i) emphasize on the current inefficiency of the three-point line from the perspective of a ``fair'' game and (ii) introduce ideas borrowed from the theory of fractals in the analysis of the game.  
There are a few things however to consider in order to place our analysis in the right context.  
We understand that pushing the three-point line 5 feet back is not trivial.  
Various other changes need to be accompanied - with the major one being the expansion of the court dimensions, especially its width.  
More importantly, the change might still not eliminate this inefficiency since players evolve and they might easily adopt to the new distance. 
Furthermore, this recommendation is based on the objective of having an (approximately) statistically constant court equity $\equity$ around the three-point line, which might as well not align with the objective in the league's eyes.  
Nevertheless, it might be appropriate for the league to test some changes in its development league and hence, get a better understanding on how such a change will impact the game. 
The development league should be seen not only as a forum for player development but also as a channel for the advancement of the game as a whole.